# A Survey of P2P Network Security

Logan Washbourne

**Abstract.** This paper presents a review of peer-to-peer network security. Popular for sharing of multimedia files, these networks carry risks and vulnerabilities relating to data integrity, spyware, adware, and unwanted files. Further attacks include those of forgery, pollution, repudiation, membership and Eclipse attacks, neighbor selection attacks, Sybil, DoS, and omission attacks. We review some protection mechanisms that have been devised.

**Introduction**
Most information transferred on the internet utilizes a server/client communication protocol, where one high performance machine services the requests of multiple lower performance machines. These lower performance machines have identities that the server can recognize, allowing a safe transaction to occur [1]. Unlike server/client networks, peer-to-peer (P2P) networks consist of nodes, or users, who behave like both server and client, able to simultaneously request information and provide information to other nodes. P2P nodes have aptly been dubbed Servents, taking half of the word server and half of the word client [1]. Not only can nodes within a P2P network both send and receive data, they can also share resources such as processing power, storage, peripherals, and network capacity [1]. P2P networks became popular after the emergence of Napster, an early application that allowed users to share music with other users. It utilized a hybrid P2P network where there was a central server that allowed users to search for music and then multiple users would contribute percentages of the file being searched for [2].

The P2P model is convenient because of its classical ability to search for files throughout the network, using a semi-centralized server; however, as it will be shown, any form of centralized node grouping can leave the whole network vulnerable to attacks. Along with hybrid P2P networks there are also "pure" P2P networks, which do not have any centralized nodes, and all file transfers happen between users.

Streaming P2P services suffer additional attacks such as data pollution that affects QoS, or data outage by the use of Eclipse, Sybil and neighbor selection attacks. This is done by attacking the routing substrate of the P2P network which is based normally on distributed hash tables [3].

The study of P2P networks is of particular interest because in many ways, in the absence of centralized control, they mimic and even amplify social behavior. The same issues that confront information integrity in knowledge networks [4],[5] show up in P2P networks. The use of random sequences for authentication protocols and discovering intruders [6]-[12] is important for the protection of social interaction. This paper presents an overview of P2P network models and considers security issues as well as applications.

**History of P2P Networking**
Peer-to-peer networking essentially started with the ARPANET, which began as a venture, through DARPA, to network together computers that were under contract with them in 1966. Larry Roberts successfully connected computers at UCLA, SRI, UCSB, and UoU, in 1969 utilizing a new method of data transfer called "packet switching [13]." The network could utilize email, file transfers, and remote logons. The ARPANET started out as an experiment to connect computers together but it was continually funded because of military interests [14].

The ARPANET allowed computers to send and receive information from other nodes on the network by the use of packet switching. Packet switching allowed information to be sent by breaking up the information into smaller parts called packets. Along with the packet a destination message was sent, telling the intermediate nodes on the network where the packets where supposed to end up. This method is analogous to P2P networking today because information can hop between nodes in order to end up at the correct destination and the network for the most part is decentralized. Version two ARPANET added the TCP/IP protocols which were used in the virtual network setup to connect multiple nodes together, regardless of their hardware configurations. The ARPANET eventually lead to the NSFnet which successfully connected 2000 university computers together. These are the antecedents to the internet of today.

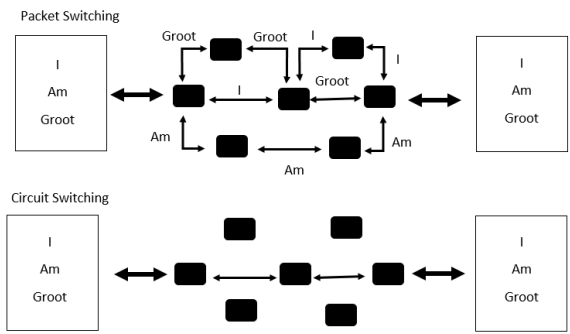

*Figure1: Packet Switching vs. Circuit Switching*

**Napster**

The first large P2P network scheme and application that the public used was Napster. It was an application that allowed users to share MP3 files with each other that was officially operational from 1999 to 2001. Napster utilized a hybrid centralized P2P model where a centralized server facilitated the pairing of two users based off of files desired and files owned closely resembling Figure 2 [15].

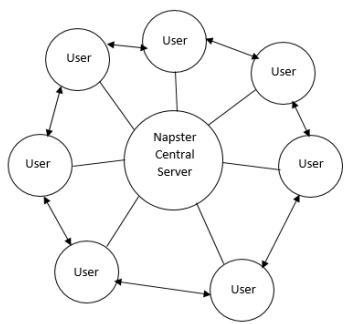

*Figure 2: Napster's centralized P2P architecture*

Napster's P2P protocol only allowed for MP3's to be indexed and searched which many considered a drawback. A work-around was developed that "wrapped" other file types into what looked like an MP3, fooling Napster's servers into being searchable. The most common tool used to wrap files was aptly dubbed "Wrapster."

When requesting file locations from the Napster server, the state of the user is relayed, making anonymity unattainable. The lack of anonymity is seen by users as a less than desirable trait for P2P networks. In order to share files, users create a TCP connection between themselves and pass files back and forth. The TCP/IP protocol has no inherent security measures, making these file transfers quite unsafe to prying eyes [16].

Another drawback to the Napster P2P network lies in its architecture. As mentioned before it revolves around a central server to facilitate all of the requests in pairing users together. This architecture is susceptible to attack and can quite easily be taken out of commission. If the central server is compromised, either by DDoS attacks or physical hardware attacks, the whole network will cease to function because the main node is inoperable.

Napster led to many clone applications that utilized P2P networking, usually with the intention of sharing files and not just MP3's. Not only were clones developed after the initial success of Napster, other P2P architecture networks were developed and adapted, hoping to create a more efficient and secure network.

**Gnutella**
Gnutella began in early 2000 at a subsidiary of America Online. The development was scrapped shortly after, but not before a few users downloaded it, leading to the reverse engineering of the application and its widespread use.

The Gnutella protocol does not rely on a central server to handle user queries; instead it utilizes a "flat ad-hoc topology [16]." Every user, or node, in the system acts as a servent, both a server and a client, able to respond to queries from neighboring nodes and issue queries to neighboring nodes as can be seen in Figure 3. The network can also be classified as a decentralized P2P network where every node is connected to many other nodes. This architecture insures the networks survivability, if one node goes offline, the whole network does not suffer.

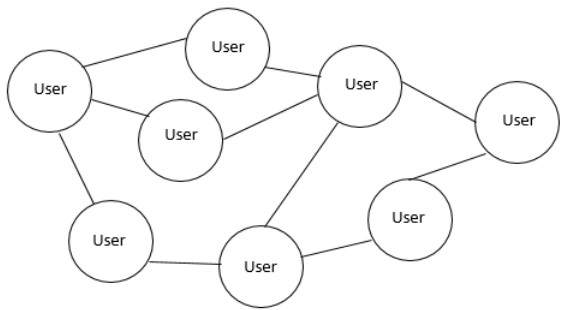

*Figure 3: Gnutella's decentralized P2P architecture*

In order for users to find files they want to download, they query their neighbors asking if they themselves have said file. If the neighbors do not have the files themselves, they in turn ask their own neighbors. This process repeats until either the file is found or the Time-To-Live (TTL) counter has decremented to zero from a certain starting number. Each node that comes across a request will decrement the TTL counter associated with that request. Once a user has found a desired file, it downloads the file directly from the user who owned it using HTTP. Just like Napster, Gnutella does not host any of the transferred files on its on servers, the file sharing happens only between users themselves.

Drawbacks of the Gnutella protocol include users who have slow transfer speeds, high bandwidth from overhead messages, and users who do not share files.

If users participate within the network and have a low transfer speed, they can fragment the network [16]. Requests from neighbors can inundate them and due to their slow response time, these nodes can essentially be crippled by normal traffic. If these nodes become crippled, they cannot respond to queries which means they will act like dead nodes, they will still be connected to their neighbors but they will not contribute to the network as a whole. This will fragment the network into pockets of "live" nodes where there will be a bottleneck to connect with other pockets.

In order to stay connected to the Gnutella network, each node needs to periodically "ping" their neighbors in order to determine if they still exist and to update their list of connected nodes. If a node receives a ping, it sends back a "pong" with information like which port is being used, the IP address of the node, number of files shared, and number of bytes shared [14]. These periodic messages add up within the network and have been estimated to contribute to roughly 50% of the network bandwidth [16]. The inherent protocol is bandwidth inefficient.

According to [17] 70% of all Gnutella users don't share files leading to 25% of the user base bearing 99% of the networks load. The protocol does not enforce users to participate either, so it's not possible to weed them out. This will lead to users not being able to find the files they are looking for and creating a very inefficient network.

Users can stay anonymous due to the "daisy-chaining" of passing requests throughout the network. If a user receives a request from a neighbor, it cannot be determined if it originated from the neighbor or further along in the network, leading itself to an anonymous action. However, every ping message contains the original user's Descriptor ID, which is a unique identifier for that user on the network. If another user can decrypt the Descriptor ID, it is possible that the original user's identity can be determined.

**BitTorrent**
The BitTorrent protocol was developed in 2001 by Bram Cohen, a University of Buffalo student at the time. BitTorrent's network architecture is much like Gnutella's, it is a decentralized P2P network where none of the file transfers occur within the protocol itself. In order for users to find other users, they must use a "tracker" which is a list of IP addresses of users sharing a certain file [17]. In order for users to download a specific file they have to find a tracker of it, which can usually be found on tracker websites like The Pirate Bay. The files you download from tracker websites include a tracker file and a torrent file. The torrent file includes the number of pieces and blocks a file contains, the IP address and port number of the tracker, and also the SHA1 hash tables of the pieces for the file. The SHA1 hash tables allow users to "verify the integrity" of each piece downloaded [18]. Files shared through a BitTorrent network have to be broken up into "pieces" and "blocks." Typically a piece is 512 kBytes and a block is 16 kBytes. When downloading files from neighboring users, the protocol allows for downloading several pieces in parallel, utilizing many neighbors at once. A user can either be a "leecher" or a "seeder" within the BitTorrent protocol. A leecher is a user who has not completely downloaded the file associated with the tracker he is a part of. A seeder is a user is has a complete version of the file and is sharing it with leechers.

In order to address the inefficiency problems that Gnutella suffered due to users who did not participate in sharing files, BitTorrent's protocol favors users who upload files. Seeders periodically check the upload rates of its neighbors and only share with those who also upload. This "chokes" users who do not upload, forcing them to have slower download speeds or even restricting them from downloading from that tracker [18].

**Security Issues**
There are many issues that can plague a P2P network and most of them are caused by the decentralized and anonymous characteristics inherent to P2P networks [19]. According to [20] the main security vulnerabilities for P2P networks are: Leechers, Social Attacks, Listening Queries, DDoS attacks, Contenet Verification, and Malware. Leechers are users who only download from other users while not sharing files or resources with others. These users contribute to an imbalance in most P2P networks where a small percentage of users bear the majority of the sharing load [21]. Having an imbalance creates an inefficient network with presumably slower transfer speeds than what could be possible.

Social attacks involve inexperienced users unintentionally divulging sensitive information about their machines to more experience P2P users who can use that information to possibly obtain their passwords or even see their whole system drive [20].

Some users act as super nodes, staying connected to the internet permanently in order to route messages and keep a list of shared files for their sub nodes. These super nodes have the potential to compile a list of queries that their sub nodes make. These lists can be used to determine a user's identity or habits, which is not the anonymous experience that P2P networks promise.

DDoS (Distributed Denial of Service) attacks aim to cripple a user or network, making it so the target cannot reply to any requests, rendering them useless. In DDoS what the attacker or attackers do is send a constant stream of "bogus" [20] packets to the target which depletes its resources, denying the target the chance to fulfill its services. DDoS attacks are not as effective on P2P networks versus server/client networks, given the decentralized nature of P2P networks. If a user is flooded by a DDoS attack, the P2P network as a whole will barely be affected because no one user is responsible for every service request, unlike a server/client network, where DDoS attacks are very effective. Of course with enough attackers targeting a majority of the users would cause severe damage to any P2P network, it is just not as effective due to the large numbers of attackers required.

One of the hardest attacks to combat is content verification. Since the network is made up of equally influential nodes, i.e. there is no central node that is verifying requests and services, there is no guarantee that users are sending each other the files they promise. This is a large contribution to the spread of malware throughout P2P networks. The biggest defense against downloading a mistitled file is to not judge the file but judge the user sending the file. In some P2P networks users can be rated for their honesty of file transfers, which allows for malicious users to be called out and to warn others not to download from them. There are a couple of models that allow P2P networks to rank their users, as seen in [21] and [22].

Rice proposes a pricing model where it is "cheaper" to download from more reputable users than it is to download from users who do not have a great rating.

**Encryption Techniques**

There are two schools of thought when it comes to P2P encryption. First is encrypting/obfuscating user traffic in order to mask the P2P protocol from ISP's or other prying eyes [23]. Second is encrypting data shared to a neighbor and only unencrypting it when that neighbor uploads that data to its neighbors [24]. This second technique is used to prevent "free-loading" users.

Initially, in order to obfuscate traffic, users would encrypt their data sent between neighbors utilizing a Diffie-Helman key exchange. The actual data being sent between two parties cannot be determined when using this type of encryption, however, the type of data can still be figured out. Only encrypting the packets will allow for users to overcome what is called deep packet inspection, or DPI, which looks at the contents of the packets being sent, although the flow of the protocol can still be determined. Every type of data traffic being sent over the application layer of the internet has its own flow, i.e. where the data packets are going, where they are coming from, packet orientation, packet size, and duration of communication. These attributes can help prying eyes determine the type of data transmission occurring between two parties. A method that can be used is called Statistical Protocol Identification, or SPID. SPID uses a statistical model to determine which protocol users are using to obtain information, like streaming, P2P, or VoIP. Users cannot only use encryption to obfuscate their traffic, they must also mask their data flow.

In [23] the authors explain the steps necessary to modify a BitTorrent client in allowing for encryption plus flow obfuscation. The first step would be to use a shared random secret in order to encrypt the packet contents which will fool any DPI systems. Next, in order to decrease the statistical likelihood of determining the BitTorrent protocol flow, the authors introduce a new message type called the padding message. Inserting these new messages into the data stream increases the packet variation, helping to counter SPID. The third alteration, random flushes, further increase packet variation and packet size, adding to the countering of SPID. Lastly, a magic peer ID, is used in order to signal to sharing neighbors that the transmission will be encrypted and obfuscated. If the sharing neighbor cannot support obfuscation then the connection is terminated and they will have to reconnect and use plain text to complete their sharing. This proposed method will not result in perfect obfuscation due to the frequency at which packets are sent is not modified and can be used to determine protocols, however, this method does help in hiding the protocol type from most methods of protocol recognition approaches.

In [22] a method called "Trick before Treating" is proposed that would decrease the number of "free-riders" within a BitTorrent network by encrypting all data that is uploaded. If the majority of users within a network are free-riders, users who do not upload any data, then the overall "capacity per user" is decreased. The Trick before Treating method requires all seeders to encrypt their file pieces when they upload them to other users. In order for those users downloading those pieces to decrypt them, they have to upload those pieces to other neighbors, who in turn give them a subkey as can be seen in Figure 4. The original seeder would also give each leecher who downloads a file piece a subkey. There would be k subkeys and a user would need n subkeys to decrypt the file pieces. In order for users to obtain the whole file and decrypt it, they would have to upload their own file pieces to other users, forcing them to contribute within the network, eliminating a large portion of free-riders.

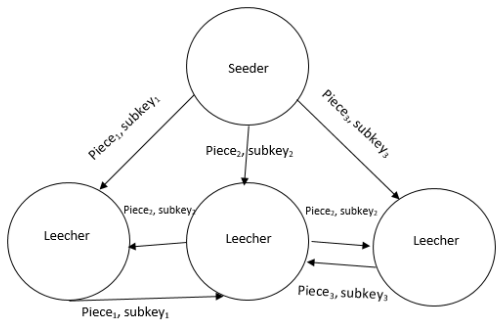

*Figure 4: Trick Before Treating*

**File Spreading Models**

In order to combat the spread of viruses and malware throughout a P2P network it must first be understood how the infected files spread through the network. Understanding how files spread can help stop the propagation of unwanted files and help improve the network structure to facilitate faster transfers. A popular model that is used is the SIR model which comes from the field of epidemiology, the study of diseases and how they spread [20].

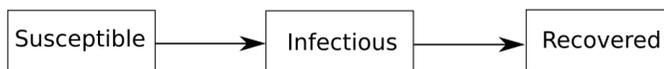

*Figure 5: SIR FileSpreading Model*

According to the SIR model, a user or person can only be in one state at a time and once they are recovered they cannot be infected again. This is a simplified model and does not work for the spread of infectious files because a system can be infected multiple times after the removal of the "diseased" file. A better model would be the modified SEIR model presented in [20].

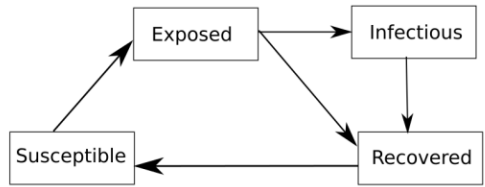

*Figure 6: SEIR File Spreading Model*

A user exposed to an infected file will not necessarily become infectious, due to virus protection and file screening. The model does point out that if a user recovers from an exposure of "infection" from a malicious file, it can become infected again, which is problematic to the P2P network that the user is connected to. The spread of malicious files is a problem to overcome in P2P networks.

**Bitcoin**
Bitcoin is a decentralized peer to peer cryptocurrency. It was proposed by Satoshi Nakamoto in 2008, in hopes to create a secure electronic cash system that would not rely on a third party for security. All transactions utilizing Bitcoin are published publicly in what is called the blockchain. The blockchain contains blocks of accepted transactions and each new block of transactions are appended to the previous blockchain, creating an ongoing public record of transactions [25]. Blocks are created by what has been dubbed as "mining." In order to mine a block, the previous block in the block chain must be hashed with SHA-256 and the transactions of the current, yet to be created, block must also be hashed. Those pieces are included in the current block along with a "Proof-of-work" identifier [26]. The proof-of-work identifier must be a nonce, a number only used once, that yields the same hash value of the current block, but the nonce must be smaller than a specified value, indicating that the user or users who discovered the nonce, had to utilize considerable CPU power.

Whenever a user mines a block and it is verified by other users, Bitcoins are awarded to him along with transaction fees from the transactions that exist within the block. This method introduces an incentive based model. Users are encouraged with compensation to preserve the transaction history of all other users, creating an incentive based decentralized cryptocurrency.

Users utilize Bitcoin wallets to make transactions with their Bitcoins. A wallet contains several addresses that point to specific Bitcoin values. Each address has a private key associated with it and the wallet stores those keys as well. Whenever a transaction is made, the public key for the Bitcoin address being used is published and can be used by any user to verify that the private key associated with the address is valid [27]. In order to send a valid transaction a digital signature must be used that reassigns the ownership of the Bitcoins within the transaction. The Bitcoin protocol utilizes the Elliptic Curve Digital Signature Algorithm to create signatures using user's private keys [28].

The following equation is the basis of Elliptic Curve Cryptography:

$$y^2 = x^3 + ax + b$$

The points that satisfy the above equation belong to the finite field based off of a chosen prime, p. Values generated by the elliptic curve equation are done so by using an initial point, G(x,y), usually designated the base point. Multiplication within ECC can be understood as adding G to itself, M times. Given the structure of ECC, the equation:

$$C = M \times G$$

Is near impossible to undo. Given M and C a user will not be able to solve for G analytically, since all values that are valid for G belong to a finite prime group. This is analogous to the Discrete Log Problem [29].

In order to utilize the ECDSA with a file or document, a user must first choose a large prime, p, an a and b parameter for the elliptic curve equation, and a base point, G. Point G must have a high order, n, which means the least number of times that G is added to itself and the result is 0 must be a large number. This number must be large in order to ensure a secure digital signature. A value Y must then be computed by multiplying a random number, X, by G. This random number must fall between 1 and n-1, and will be called the user's secret key. The Y value computed is called the user's public key. Values p, a, b, G, n, and Y are all public knowledge. The user then must hash their file, typically with SHA-256, and then choose a random value, k, between 0 and n-1. This k value will be used to create the digital signature, however, each time a digital signature is created for a file a new k value must be chosen or the robustness for the digital signature drops significantly [30]. Lastly the user creates and sends these following values to whoever is requesting their file along with the document itself;

$$Sig_1 = (K \times G)_x \bmod n$$

$$Sig_2 = K^{-1}(M + X * sig_1) \bmod n$$

The subscript x within the Sig1 equation denotes the x coordinate that yields from that operation. Every time a transaction is sent to other user's to validate the above ECSDA must be used. This algorithm must be secure otherwise users could claim transactions as their own or reroute transactions to other places, invalidating the Bitcoin protocol. The algorithm cannot be broken analytically because of the discrete log problem, but the brute force method must be investigated.

In order to manipulate a transaction an eavesdropper would have to solve for X and K. For the case of X, the eavesdropper already knows that

$$Y = G \times X$$

of which Y and G are known. The eavesdropper will have to test value of X that when multiplied by G = Y, however, X can be any number between 1 and n-1, and n can be anywhere between 56 bits to 256 bits. To be consistent with SHA-256 we will say that n is between 1 and 256 bits, or $2^{256}$, which is larger than $1*10^{77}$. Conservatively, a computer can test a value of X in 1 processor instruction, in 2014 a consumer i7 processor could execute 298,190 million instruction per second. That would mean this computer would take $3.67*10^{69}$ years to break determine the value of X. The brute force method is too slow for ECC when large key spaces are used.

**Conclusions**

Within this paper, the history of P2P networks is discussed along with security flaws, encryption techniques and file spreading models. The history of emerging P2P protocols is rich and the motivation behind developing them has changed over time. Beginning with connecting researchers and pushing the bounds of computation to citizens sharing pirated media and currently revolutionizing digital currency around the world. P2P networks have made an immense impact on the world in which we operate today. But this success comes with risks and vulnerabilities relating to data integrity, malware, and unwanted files. Research in this area must not only devise new protections for emerging threats but also deal with the objective of anonymity and forensics [31]-[39].

Trust is an important notion in data access and dissemination. As new applications of distributed systems have emerged, trust relationships among system users and service providers have become tangled. One would increasingly need high level of cross-boundary capability of trust management for local policies and credentials are not sufficient to evaluate mutual trustworthiness.

30. Kak, A. Computer and Network Security. Purdue University, 2015. https://engineering.purdue.edu/kak/compsec/NewLectures/Lecture14.pdf
31. Kim, J. Security issues in peer to peer systems. Advanced Communication Technology, The 7th International Conference of Advanced Communication Technology (ICACT 2005 . 11 July 2005, p. 1059.
32. Bailes, J., Templeton, G.. Managing P2P security. Communications of the ACM 47 (9): 95 – 98 (2004)
33. McDowell, M., Wrisley, B., Dormann, W.. Risks of file sharing technology. National Cyber Alert System, Cyber Security Tip ST05-007, Carnegie Mellon University. May 19, 2010.
34. Chen, R., Lua, E.K., Crowcroft, J.. Securing peer-to-peer content sharing service from poisoning attacks. P2P '08 Proceedings of the 2008 Eighth International Conference on Peer-to-Peer Computing. IEEE Computer Society, pp 22-29.
35. Barcellos, M. P2P-SEC security issues and perspectives on P2P systems: from Gnutella to BitTorrent. 53nd International Federation for Information Processing (IFIP) 10.4 Working Group on Dependable Computing and Fault Tolerance, Feb 2008.
36. Tanenbaum, A., Van Steen, M. Distributed Systems: Principles and Paradigms, 2nd Ed. Prentice-Hall 2007.
37. Chan-Tin, E., Chen, T., Kak, S. A comprehensive security model for networking applications. MobiPST, Munich, July 30- Aug 2, 2012.
38. Kak, S., The piggy bank cryptographic trope. Infocommunications Journal, vol. 6, pp. 22-25, March 2014.
39. Mhapasekar, D. Accomplishing anonymity in a peer to peer network. Proceedings of the 2011 International Conference on Comm, Computing & Security. February 12-14, 2011.
12